\documentclass[%
 reprint,
 amsmath,amssymb,
 aps,prl
]{revtex4-2}

\usepackage{graphicx}

\usepackage{bm}
\usepackage{siunitx}
\usepackage{braket}

\usepackage{tikz}
\usepackage{hyperref}

\begin{document}


\title{Thermal melting of a vortex lattice in a quasi two-dimensional Bose gas}

\author{Rishabh Sharma}
\author{David Rey}
\author{Laurent Longchambon}
\author{Aurélien Perrin}
\author{Hélène Perrin}
\author{Romain Dubessy}
\email{romain.dubessy@univ-paris13.fr}
\affiliation{Université Sorbonne Paris Nord, Laboratoire de Physique des Lasers, CNRS UMR 7538, 99 av. J.-B. Clément, F‐93430 Villetaneuse, France}

\date{\today}

\begin{abstract}
We report the observation of the melting of a vortex lattice in a fast rotating quasi-two dimensional Bose gas, under the influence of thermal fluctuations. We image the vortex lattice after a time-of-flight expansion, for increasing rotation frequency at constant atom number and temperature. We detect the vortex positions and study the order of the lattice using the pair correlation function and the orientational correlation function. We evidence the melting transition by a change in the decay of orientational correlations, associated to a proliferation of dislocations. Our findings are consistent with the hexatic to liquid transition in the KTHNY scenario for two-dimensional melting.
\end{abstract}

\maketitle


The dimensionality in which a quantum system evolves plays a crucial role for its properties, increasing the role of fluctuations and challenging the establishment of coherence~\cite{Bloch2008RMP}. While true long-range ordered crystals exist in three dimensions, the Mermin-Wagner-Hohenberg theorem prevents the existence of long-range order in one or two dimensions~\cite{Mermin1966,Hohenberg1967}. However, for a sufficiently low temperature, a quasi-long range order is expected and a two-stage scenario for the melting of a two-dimensional (2D) crystal has been  proposed by Kosterlitz, Thouless, Halperin, Nelson and Young~\cite{Kosterlitz1973,Halperin1978,Young1979} (KTHNY), involving the thermal activation and unbinding of lattice defects. This melting scenario of  2D crystals has attracted a lot of attention for classical as well as for quantum systems.

The nature of the transition depends on the interaction potential, and there has been an extensive numerical~\cite{Iaconis2010,Bernard2011,Wierschem2011,Kapfer2015,Li2019} and experimental~\cite{Murray1987,Tang1989,Kusner1994, Marcus1996,Zahn1999,Han2008,Gasser2010,Kelleher2017,Petrov2015,Sun2016,Guillamon2009,Guillamon2014,Zehetmayer2015,Roy2019,Huang2020} activity to define the validity of this theory. Experiments are based mostly on a large number of small particles such as colloidal solutions~\cite{Murray1987,Tang1989,Kusner1994,Marcus1996,Zahn1999,Han2008,Gasser2010,Kelleher2017}, and air-fluidized dust or spheres~\cite{Petrov2015,Sun2016}. Whereas these systems are made of classical objects, the melting of crystals according to the KTHNY scenario has also been observed in the case of quantum vortex lattices in thin superconductors~\cite{Guillamon2009,Guillamon2014,Zehetmayer2015,Roy2019} and for a lattice of skyrmions~\cite{Huang2020}. A fast rotating superfluid also exhibits a large number of quantized vortices  that arrange in a triangular lattice, as observed in superfluid helium~\cite{Yarmchuk1979} and in dilute Bose-Einstein condensates (BEC)~\cite{Abo-Shaeer2001,Coddington2003,Bretin2004}.

The specific case of vortex lattice melting in a rotating BEC confined in a rotationally invariant harmonic trap has been discussed by Gifford and Baym~\cite{Gifford2008} on the basis of the KTHNY scenario. They give an evaluation of a melting temperature $T_m$, at fixed rotation frequency $\Omega$ and for a given trap geometry. Interestingly, the estimated melting temperature depends on the trap anisotropy $\omega_r/\omega_z$, the ratio between the radial and longitudinal trapping frequencies. For cigar-shaped traps where $\omega_r>\omega_z$, as is the case in Ref.~\cite{Abo-Shaeer2001,Coddington2003,Bretin2004}, the melting temperature is close to the condensation temperature, such that samples with a large superfluid fraction always belong to the crystalline phase. This is favorable for the observation of low energy excitations such as Tkachenko modes~\cite{Schweikhard2004}, while observing the transition to a vortex liquid phase is challenging~\cite{Gifford2008}. On the other hand, $T_m$ is expected to be significantly smaller than the superfluid critical temperature for 
oblate gases with $\omega_r<\omega_z$, giving room for the observation of the melting transition deeply in the superfluid phase. Vortex lattices in quasi-2D Bose gases are thus particularly appealing for the study of the KTHNY scenario of thermal melting.

\begin{figure}[t]
    \centering
    \includegraphics[width=8.6cm]{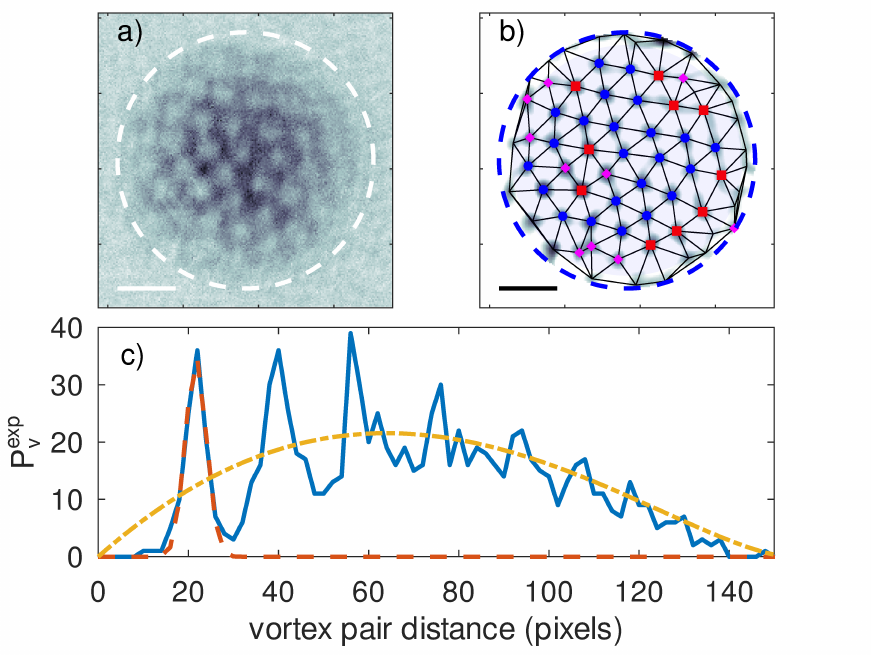}
    \caption{(color online) a) Integrated time-of-flight density profile of a quasi-two-dimensional rotating Bose gas with vortices, b) automatic detection of the vortex cores and Delaunay triangulation and c) distribution of the vortex pair distances $P_v^{\rm exp}$. In a) and b) the dashed circle indicates the Thomas-Fermi radius $R_{\rm TF}$ and the horizontal line is a 40 pixels reference scale. In b) the blue disks, magenta diamonds and red squares show lattice sites with 6, 5 and 7 neighbors, respectively. In c) the red dashed line is a Gaussian fit to the first peak. The dashed-dotted yellow curve is the prediction $\,\overline{\!P_v}$ for a uniform continuous vortex distribution. See text for details.}
    \label{fig:1}
\end{figure}

In this Letter, we present the first observation of the thermal melting of a vortex lattice in a rotating quasi-2D superfluid Bose gas. For experimental convenience, the parameter tuned to induce the melting is not the gas temperature $T$ but instead its rotation rate $\Omega$ at fixed temperature. From the density distribution of the superfluid imaged after a time-of-flight expansion, we reconstruct the positions of the vortices and identify their nearest neighbors as shown in Fig.~\ref{fig:1}a) and b). We characterize the vortex lattice translational order with the pair correlation function $g(r)$ and its orientational order with the orientational correlation function $G_6(r)$ defined below. We show that both functions display an evolution of their features with increasing $\Omega$ that evidences a transition from a hexatic to a liquid phase. We compare the transition threshold to the predictions of Gifford and Baym~\cite{Gifford2008}. We also show that the loss of orientational order and the probability of finding lattice defects display a threshold with $\Omega$ consistent with the KTHNY scenario.

In the KTHNY scenario, two threshold temperatures $T_1<T_2$ are involved. Below $T_1$, the lattice exhibits a quasi-long range order, with possible defects consisting in bound pairs of dislocations. Dislocations are composite defects made of two adjacent sites with 5 and 7 nearest neighbours. Ref.~\cite{Gifford2008} estimates an upper bound $T_m$ for $T_1$. Above $T_1$, thermal fluctuations activate the unbinding of the dislocation pairs. The proliferation of free dislocations leads to the loss of the quasi-long range translational order, while maintaining a quasi-long range orientational order, defining the hexatic phase~\cite{Nelson1977}. Then, above the second threshold $T_2$, dislocations turn into free 5-fold and 7-fold defects, which destroys the orientational order, and the system enters the liquid phase.

The orientational order is characterized by the orientational correlation function $G_6(r)$~\cite{Halperin1978}. It is defined from the orientation order parameter, given for each vortex $k$ at position $\bm{r}_k$ by: $\Psi^{(6)}_k=\left(\sum_{j=1}^{N_k}e^{6i\theta_{kj}}\right)/N_k$, where the sum runs over the $N_k$ first neighbors and $\theta_{kj}$ is the angle between a fixed axis and the bond linking the sites $k$ and $j$. The orientational correlation function is then
\begin{equation}
    G_6(r)=\braket{\Psi^{(6)*}_k\Psi^{(6)}_p}_{|\bm{r}_k-\bm{r}_p|\sim r} \label{eq:G6}
\end{equation}
where the average is taken over all pairs of vortices split by a distance between $r-\delta r/2$ and $r+\delta r/2$~\footnote{We use $\delta r=2$ pixels. With our imaging system the effective size of one pixel is $\SI{1.03}{\micro\metre}\times\SI{1.03}{\micro\metre}$.}. The KTHNY theory predicts a transition from an algebraic to an exponential decay of $|G_6(r)|$ at the transition from the hexatic to the liquid phase.

To explore this transition for quantized vortices in a superfluid, we prepare vortex lattices in rotating quasi-2D Bose gases. The experimental setup is described in~\cite{Guo2020}. Briefly, we use the radio-frequency (rf) dressed adiabatic potentials technique~\cite{Garraway2016,Perrin2017} to create an ellipsoidal shell trap, rotationally invariant around the vertical axis. In the presence of gravity, atoms gather at the bottom of this shell, resulting in an oblate trap, well described by a vertical harmonic potential of frequency $\omega_z=2\pi\times\SI{360}{\hertz}$ and a radial potential of harmonic frequency $\omega_r=2\pi\times\SI{34}{\hertz}$ with a small quartic anharmonicity~\cite{SM}. We load this trap with a degenerate Bose gas of approximately $\num{2.5e5}$ $^{87}$Rb atoms.
The vortex lattice is obtained after rotating an elliptical deformation of the trap for 10 turns, after which we let the gas relax for \SI{12}{s} in the initial rotationally invariant trap~\cite{Guo2020}. The final effective rotation rate $\Omega$ of the superfluid is tuned by adjusting both the rotating rate and the ellipticity of the deformation. The density profile integrated along the vertical axis is recorded after a time-of-flight expansion with standard absorption imaging. Figure~\ref{fig:1}a) shows a typical image of a quasi-2D rotating Bose gas released with a visible vortex lattice. We calibrate the rotation rate $\Omega$ from a comparison of the measured Thomas-Fermi (TF) radius $R_{\rm TF}$ with a numerical simulation~\cite{SM}.

\begin{figure}[t]
\centering
\includegraphics[width=8.6cm]{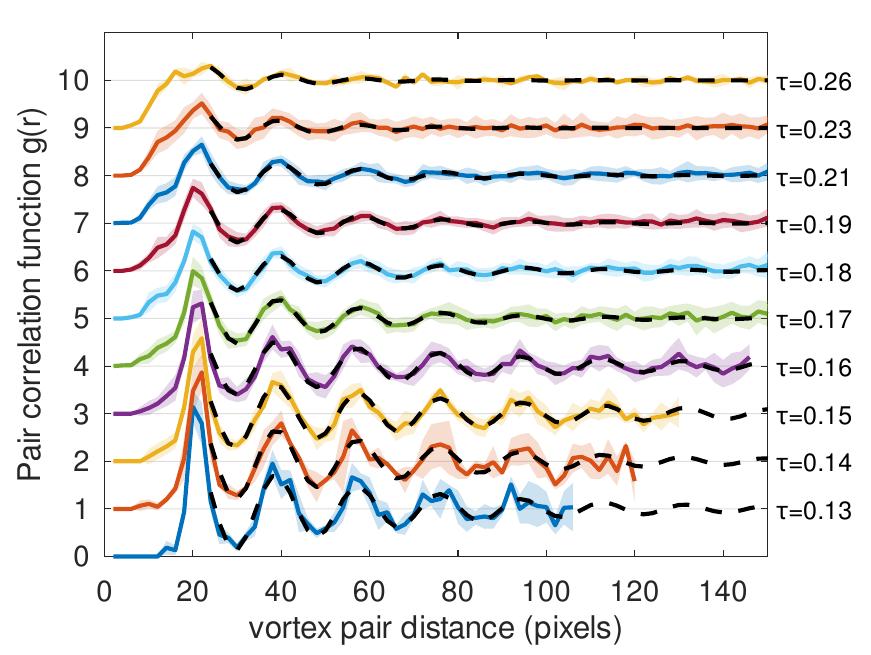}
\caption{
(color online) Pair correlation function $g(r)$ evidencing a loss of order for increasing reduced rotation frequency $\Omega/\omega_r=\{$0.70,0.75,0.80,0.85,0.875,0.90,0.925,0.95,0.975,1.0$\}$, from bottom to top. The black dashed curves show the fit of the decaying oscillations. Each curve is shifted vertically by 1 for clarity, with the reduced temperature $\tau$ indicated on the right-hand side.
The shaded areas indicate the one-sigma statistical uncertainty arising from several realizations.
}
\label{fig:2}
\end{figure}

We repeat the experiment for various values of the stirring frequency and trap anisotropy, corresponding to effective reduced rotation frequencies $\Omega/\omega_r$ in the range $(0.5,1)$. Reaching $\Omega\sim\omega_r$ is made possible by the small radial quartic anharmonicity~\cite{Bretin2004,Guo2020}. The final temperature remains fixed at $T=\SI{18}{nK}$, imposed by a rf knife, and we post select the realizations with a total atom number $N=10^5\pm10\%$. From the known trap geometry~\cite{SM} we estimate the quasi-2D critical temperature for the superfluid transition $T_{\rm BKT}(N,\Omega)$ with a semi-classical approach~\cite{Holzmann2008} and label each realization of the experiment by $\tau=T/T_{\rm BKT}(N,\Omega)$. For all experiments reported here, $\tau<0.3$ and the gas is deeply in the superfluid phase, with a negligible thermal fraction. 

\begin{figure*}[t]
\centering
\includegraphics[width=8.6cm]{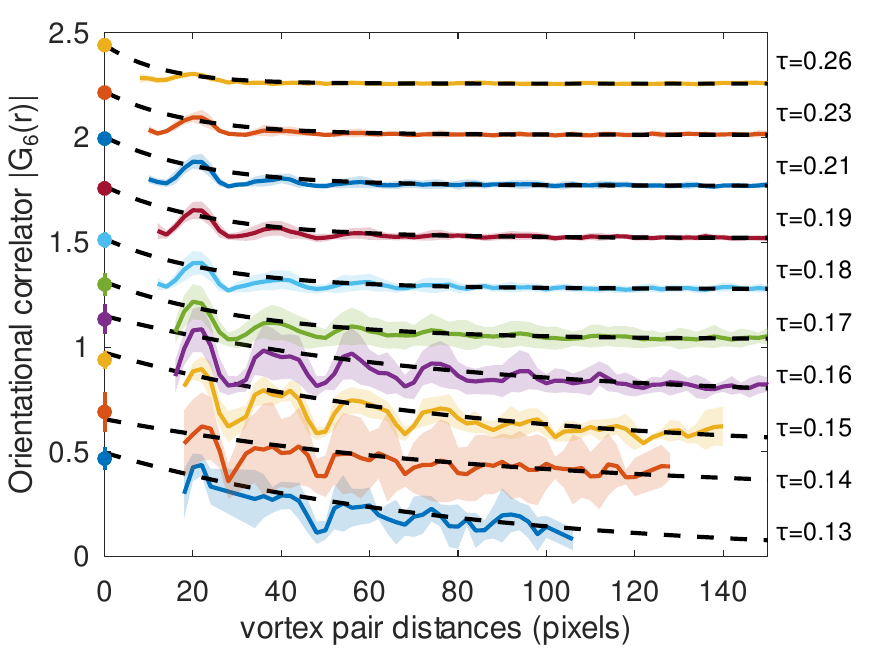}
\includegraphics[width=8.6cm]{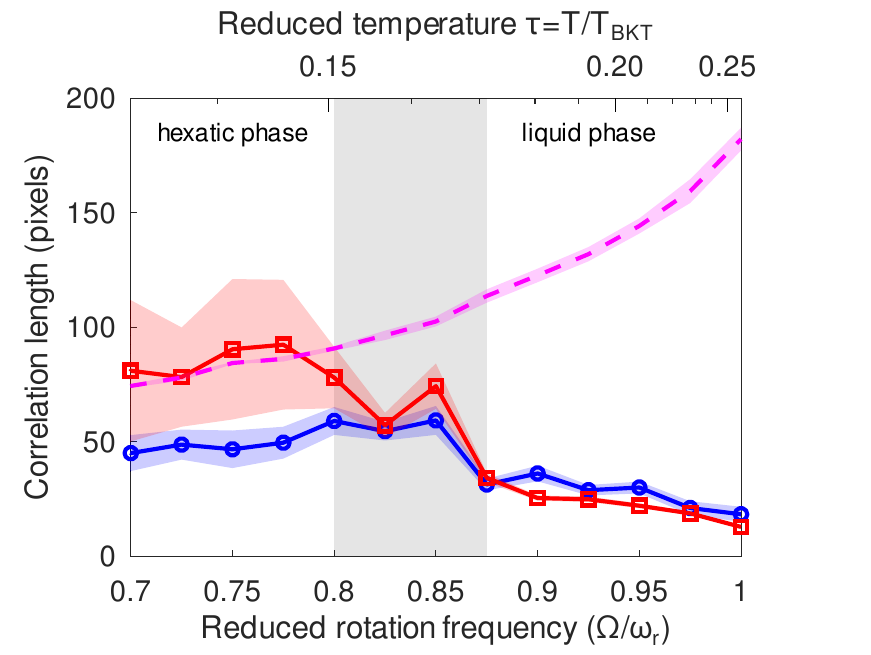}
\caption{
(color online) a) Decay of the orientational correlations $|G_6(r)|$ with increasing reduced
rotation frequency, for the same values as in Fig.~\ref{fig:2}. Each curve is shifted vertically by $0.25$ for clarity, with the reduced temperature $\tau$ indicated on the right-hand side. The black dashed curves are fits of the envelopes by decaying exponential models. The shaded areas indicate the one-sigma statistical uncertainty arising from several realizations.
b) Correlation lengths $\ell_G$ and $\ell_g$ extracted from the fitted exponential decaying envelope of $|G_6(r)|$ (red open squares) and $g(r)$ (blue open circles), respectively, as a function of the 
reduced
rotation frequency (bottom axis) or reduced temperature (top axis, notice the nonlinear scale). The solid lines connecting the points are guides to the eye. The light blue and red shaded area indicate the uncertainty on the values extracted from the fit. The dashed magenta line shows the TF radius for each reduced rotation frequency, which gives the system size. The shaded grey region indicates the crossover between the hexatic and liquid phases. See text for details.
}
\label{fig:3}
\end{figure*}

The position of the vortex centers is detected automatically from the density profile with a method adapted from Ref.~\cite{Rakonjac2016}: we first apply a Gaussian filtering to remove imaging noise, then compute the Laplacian of the image and find all local maxima, corresponding to dips in the initial density profile~\cite{SM}. Fig.~\ref{fig:1}b) shows the result of the vortex detection algorithm. Once the vortices are identified, we perform a Delaunay triangulation to define the elementary cells of the lattice. We may then label each vortex by its number of neighbors in the lattice, to identify the defects. Because of the low signal to noise ratio near the edge of the cloud we only consider the vortices in a disk of radius $R=0.9\times R_{\rm TF}$.

To characterize the order of the lattice we first study the histogram of vortex pair distances, the number $P_v^{\rm exp}(r)$ of pairs of vortices split by a distance between $r-\delta r/2$ and $r+\delta r/2$~\cite{Note1}, which displays peaks corresponding to long range order in the lattice, see Fig.~\ref{fig:1}c). By fitting the position of the first peak with a Gaussian we extract the mean next-neighbor distance $a$ that we convert into a mean vortex density $n_v=2/(\sqrt{3}a^2)$.
The peaks sit on a background well captured by the dashed-dotted yellow line, related to the finite size of the cloud. This background, which spreads between 0 and $2R$, corresponds to a similar histogram of pair distances, computed in a disk of radius $R$ for a continuous uniform vortex density $n_v$~\cite{SM}: 
\begin{equation}
\,\overline{\!P_v}(r)=2\pi n_v^2R^2r\,\delta r\left(
\arccos{\left[\frac{r}{2R}\right]}
-\frac{r}{2R}\sqrt{1-\frac{r^2}{4R^2}}
\right).
\label{eqn:gref}
\end{equation}

Figure~\ref{fig:2} shows the evolution of the pair correlation function $g(r)=P_v^{\rm exp}/\,\overline{\!P_v}$, computed as the ratio between the experimental histogram $P_v^{\rm exp}(r)$ and the analytical formula $\,\overline{\!P_v}(r)$ of Eq.~\eqref{eqn:gref}, for increasing $\Omega$~\footnote{We have to adjust slightly the vortex density $n_v$, by up to $24\%$ for the largest values of $\Omega$, to reach $g(r)\sim1$ at large distances.}. The curves are obtained by averaging at least three experimental realizations, corresponding to the same rotation frequency within $\pm1.25\%$. The data evidence several features of a generic solid to liquid transition: at low reduced temperature $\tau$, $g(r)$ displays oscillations characteristic of long range order in the sample, while the amplitude of the first peak decreases and the oscillations disappear as $\tau$ increases. We fit the decay of the oscillations by $f_g(r)=1+A_g\cos{[kr+\delta]}e^{-r/\ell_g}$ to extract a correlation length $\ell_g$~\cite{Wierschem2011}, reflecting the translational order.

\begin{figure}[t]
\centering
\includegraphics[width=8.6cm]{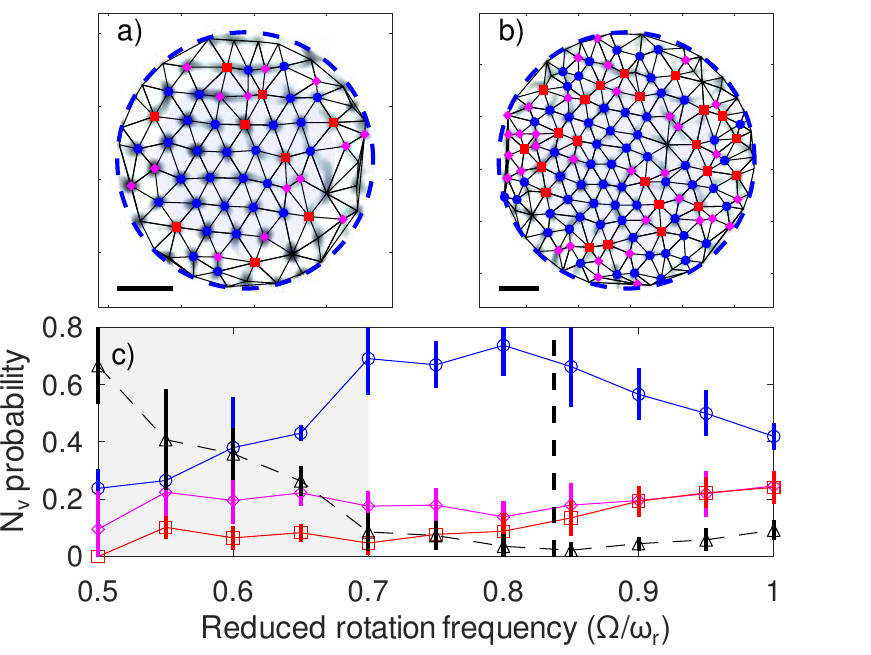}
\caption{
(color online) a) Vortex lattice for $\Omega/\omega_r=0.76$, b) vortex lattice for $\Omega/\omega_r=0.9$. Blue disks, magenta diamonds and red squares indicate lattice sites with 6, 5 and 7 neighbors, respectively. The horizontal black line is 40 pixels long. c) Probability to find a vortex with $N_v$ neighbors, as a function of $\Omega/\omega_r$: blue circles $N_v=6$, magenta diamonds $N_v=5$, red squares $N_v=7$ and black triangles all other values. The thick dashed vertical black line indicates $\Omega_m\simeq0.84\times\omega_r$. The error bars correspond to the one-sigma statistical uncertainty. Finite size effects are dominant in the light grey shaded area. See text for details.
}
\label{fig:4}
\end{figure}

We now investigate the orientational order and compute the orientational correlation function $G_6(r)$, Eq.~\eqref{eq:G6}, for each experimental realization. The nearest neighbors are deduced from the Delaunay triangulation as shown in Fig.~\ref{fig:1}b).
Figure~\ref{fig:3}a) shows $|G_6(r)|$ as a function of the vortex pair distance for increasing rotation frequencies, evidencing that the decay occurs on a shorter length scale as $\Omega$ increases. We fit the decaying envelope by an exponential model $f_e(r)=A_G\,e^{-r/\ell_G}+B$ and extract a correlation length $\ell_G$.

Figure~\ref{fig:3}b) compares the two correlation lengths $\ell_g$ and $\ell_G$ as we increase the rotation frequency. We find that $\ell_g$ is always significantly smaller than the typical system size, given by the TF radius and therefore there is no long range translational order in our samples. On the contrary, $\ell_G$ is of the order of $R_{\rm TF}$ at low rotation frequency, $\Omega\leq0.8\times\omega_r$, and drops significantly for $\Omega>0.875\times\omega_r$. This shows that at low rotation frequency the orientational order extends on the size of the system while translational order is already lost. This is a signature that the vortex lattice is in the hexatic phase and enters the liquid phase at higher rotation frequency. From our measurements we identify a reduced melting temperature of $\tau_m=0.16\pm0.01$ (corresponding to $\Omega_m/\omega_r=0.84\pm0.04$), significantly smaller than the upper bound $\tau\simeq0.5$ predicted by Gifford and Baym~\cite{Gifford2008,SM}.

\paragraph{Discussion} One remarkable feature of Fig.~\ref{fig:2} is that the position of the first neighbor peak seems to be independent of the rotation frequency $\Omega$, leading to an \emph{apparent} constant vortex density $n_v$. Using classical field simulations~\cite{Bradley2007} we have checked that this is a consequence of the finite time-of-flight expansion~\cite{SM}: the \emph{in situ} vortex density is equal to $M\Omega/(\pi\hbar)$, as expected.

While a fast rotating quasi-2D Bose gas, in a potential virtually free of any defects that could act as a pinning potential for the lattice, offers an ideal platform to study the KTHNY scenario, the main limitation of this system comes from finite size effects: our data do not clearly show a transition from an algebraic to an exponential decay~\cite{SM}; and the circular boundary of the condensate itself induces defects in the lattice. As the density decreases with the radial coordinate, one can also expect that the melting occurs first at the edge of the lattice, possibly smearing out the transition.

In order to quantify the impact of finite size effects we have studied the distribution of defects in the vortex lattice, within a radius of $R=0.7\times R_{\rm TF}$, thus avoiding the low density region at the edge of the cloud. Fig.~\ref{fig:4}a) and b) show two typical lattices below and above the hexatic to liquid phase melting transition. Fig.~\ref{fig:4}c) displays the probability for a site to have a given number of neighbors as a function of the rotation frequency. For $\Omega/\omega_r<0.7$ we find that a significant fraction of sites have less than 5 or more than 7 neighbors. We attribute this to the frustration of small lattices by the circular boundary.

We also observe that the probability of finding sites with 5 neighbors is larger than in the case of 7 neighbors for $\Omega/\omega_r<0.9$. We explain this by vortex detection errors when the algorithm artificially counts a single vortex twice with nearby positions, which tends to favor 5-neighbor sites. However the continuous increase of the probability of finding sites with 7 neighbors is associated to the proliferation of 5-7 dislocations, either bound or unbound, and to a significant decrease of the number of sites with 6 neighbors for $\Omega>\Omega_m$. This observation also supports a melting according to the KTHNY scenario.

Finally we emphasize that we observe the melting of the lattice at a temperature significantly lower than the upper bound given by the theoretical prediction~\cite{Gifford2008}. Several factors may explain this discrepancy: firstly, the system is not strictly 2D and the models should include quasi-2D corrections, accounting for particles in the first few vertical excited states; secondly, the density is inhomogeneous such that we rely on local density approximation to compare to the KTHNY predictions; then, the vortex lattices have a finite extension and the melting scenario can be affected by the boundaries; finally, the vortex lattice order may be altered by the time-of-flight expansion. The first two effects are usually well controlled in dilute atomic superfluids and should not change significantly the theoretical predictions. To study finite size effects it would be relevant to repeat the experiment with larger lattices or at fixed rotation frequency by changing the temperature, which is challenging in our system. Instead, an option would be to investigate the finite-size effects on the melting transition using intensive classical field simulations. Finally, preliminary classical field simulations suggest that the lattice deformation during expansion is weak~\cite{SM}. A direct way to eliminate this effect would be to implement \emph{in situ} vortex imaging schemes~\cite{Wilson2015}, with a large field of view.

\paragraph{Conclusion} We have evidenced the thermal melting of a vortex lattice in a quasi-2D fast rotating Bose gas, characterized in terms of the pair correlation function $g(r)$ and of the orientational correlation function $G_6(r)$. We have identified for the first time the hexatic and liquid phases in superfluid vortex lattices. We observe a transition from a slow to a fast decay of $G_6(r)$ as we cross the melting point, associated with a proliferation of dislocations above the melting point. This work paves the way for the study of thermal melting in the $\Omega\sim\omega_r$ limit, where a description in terms of lowest Landau levels becomes relevant. Decreasing the atom number and the temperature would allow us to investigate how the overlap between vortex cores affects the KTHNY scenario.

\begin{acknowledgments}
LPL is UMR 7538 of CNRS and Sorbonne Paris Nord University.
We acknowledge experimental support from Thomas Badr and Matthieu de Goër de Herve at an early stage of this project.
This work has been supported by Région Île-de-France in the framework of DIM SIRTEQ (project Hydrolive).
We acknowledge financial support from the ANR project VORTECS (Grant No. ANR-22-CE30-0011).
\end{acknowledgments}

\appendix


%

\clearpage

\renewcommand\thefigure{S\arabic{figure}}    
\setcounter{figure}{0}
\centerline{\textbf{Supplementary Material}}

\section{Experimental details}
We prepare a degenerate Bose gas of rubidium 87 atoms in a shell-shaped dressed quadrupole trap \cite{Merloti2013,Guo2020}. The horizontal quadrupole trap gradient in units of frequency is $\alpha=2\pi\times\SI{3.93}{kHz/\micro\metre}$, the rf dressing frequency $\omega=2\pi\times\SI{300}{kHz}$ and the rf coupling $\Omega_0=2\pi\times\SI{49}{\kilo\hertz}$. The cloud sits at the bottom of the shell and is highly oblate, with oscillation frequencies in the harmonic limit $\omega_r=2\pi\times\SI{34}{\hertz}$ and $\omega_z=2\pi\times\SI{360}{\hertz}$. The rf field is generated by three centimeter-scale coils with orthogonal axes, fed by a multiple-output custom DDS source and the polarization is optimized to be as close as possible to circular with respect to the $z$ axis. This ensures that the trap is rotationally invariant. A fourth output of the rf source is used to generate another rf field of initial frequency $\omega_{\rm kn}=2\pi\times\SI{80}{kHz}$, that we use as an rf-knife to control the temperature.

To set the atomic cloud into rotation we use the following procedure. We first change slightly the rf polarization to introduce a small in-plane anisotropy: $\omega_x=\omega_r\sqrt{1+\epsilon}$ and $\omega_y=\omega_r\sqrt{1-\epsilon}$ with $\epsilon\in(0.04,0.2)$, then rotate its axis at a frequency $\Omega_{\rm stir}$ for $10$ full revolutions with $\Omega_{\rm stir}\in(0.59,0.88)\times\omega_r$, wait for $\SI{1}{s}$, lower the rf-knife to $\omega_{\rm kn}=2\pi\times\SI{60}{kHz}$ in $\SI{1}{s}$ and wait for $\SI{8}{s}$. The effective rotation frequency $\Omega$ of the atomic cloud is deduced from the analysis of the density profile, see below, and is determined by the interplay of the excitation procedure and a spin-up evaporation due to the rf-knife~\cite{Guo2020}. We obtain a similar atom number $N=10^5$, within \SI{10}{\percent}, for various final effective rotation frequencies $\Omega$.

In the rotating frame the atoms tend to explore a larger fraction of the rf-dressed shell trap. For the rotation frequencies considered in this work we can model the trap potential as a harmonic oscillator with a quartic correction in the horizontal plane~\cite{Guo2020}:
\begin{equation}
    V(r,z)=\frac{M\omega_r^2r^2}{2}\left(1+\kappa\frac{r^2}{a_r^2}\right)+\frac{M\omega_z^2z^2}{2},
    \label{eq:potential}
\end{equation}
where $\kappa\simeq\SI{1.5e-4}{}$ and $a_r=\sqrt{\hbar/M\omega_r}$. The effective potential in the presence of rotation is modified by the centrifugal potential, $V_\Omega(r,z) = V(r,z) - M\Omega^2r^2/2$. The equilibrium properties of a superfluid in this type of potential are well known~\cite{Cozzini2005}.

Finally we image the rotating gas after a \SI{27}{ms} time-of-flight expansion, following an abrupt switch off of the quadrupole coils currents and rf fields. Just before the switch-off the rf dressing frequency $\omega$ is ramped to $\omega^\prime/(2\pi)=\SI{340}{kHz}$ in $\SI{300}{\micro\second}$ to transfer the atomic cloud from the dressed state $m_F^\prime=1$ to the bare spin state $m_F=-1$ to avoid spin separation during the time-of-flight expansion that would blur the vortex lattice. The cloud is imaged with a standard absorption imaging scheme, with a magnification of $7.73$ and a resolution of $\sigma=\SI{4}{\micro\metre}$, resulting in a time-of-flight density profile integrated along the rotation axis, with usually many visible vortices. The expansion is essential to magnify the vortex core size and make it visible given the optical resolution.

\section{Uniform vortex distribution}
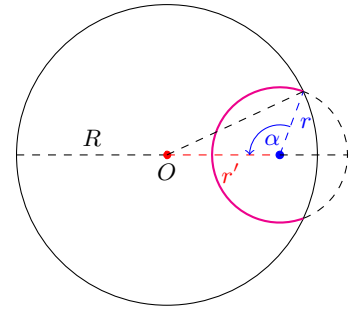
\begin{figure}[t]
\centering
\begin{tikzpicture}
\draw (2,0) arc (0:360:2);
\draw[dashed] (0,0) node[below] {$O$} --node[above] {$R$} (-2,0);
\draw[red,dashed] (0,0) node[inner sep=1,fill,circle,draw] {}--node[below,xshift=1mm] {$r^\prime$} (1.5,0);
\node[inner sep=1,fill,circle,draw,blue] at (1.5,0) {};
\draw[dashed] (1.5,0)++(-69.63:0.9) arc (-69.63:69.63:0.9);
\draw[magenta,thick] (1.5,0)++(69.63:0.9) arc (69.63:290.37:0.9);
\draw[blue,dashed] (1.5,0) --node[right] {$r$} ++(69.63:0.9);
\draw[blue,->] (1.5,0)++(69.63:0.4) arc (69.63:180:0.4) node[xshift=3mm,yshift=2mm] {\small $\alpha$};
\draw[dashed] (0,0) -- ++(24.95:2);
\draw[dashed] (1.5,0) -- (2.4,0); 
\end{tikzpicture}
\caption{(color online) Simple sketch to illustrate the problem of counting how many vortex pairs are to be found within a given distance of each other. Assuming a uniform vortex density over a disk it is equivalent to finding the length of the magenta arc, see text for details.}
\label{fig:gref}
\end{figure}

Here we derive Eq.~(2) 
of the main text. We want to compute the number of pairs split by a distance $r\pm\delta r/2$, assuming a uniform continuous density $n_v$ in a disk of radius $R$. Considering a point at a distance $r^\prime$ from the center, the number of points inside the disk at a distance $r$ is given by $n_v\times r\,\delta r\times2\alpha$, where the angle $\alpha$ is defined on the sketch of Fig.~\ref{fig:gref}. This angle, between $0$ and $\pi$, is uniquely defined by the relations:
\[
\begin{array}{l}
\cos{\alpha}=\displaystyle\frac{r^2+r^{\prime2}-R^2}{2rr^\prime}\quad\mbox{for }r^\prime>|R-r|,\\[2mm]
\alpha=0 \quad\mbox{for } r^\prime<r-R, \quad \mbox{with } R<r,\\
\alpha=\pi \quad\mbox{for } r^\prime<R-r, \quad\mbox{with } R>r.
\end{array}
\]

Finally we have to average over all initial values for $r^\prime$ inside the disk:
\[
\,\overline{\!P_v}=2\pi n_v\int_0^R r^\prime dr^\prime\,n_v\, r \,\delta r\times 2\alpha,
\]
which gives the result of Eq.~\eqref{eqn:gref} after straightforward integration.

\section{Estimation of the melting temperature}
In Ref.~\cite{Gifford2008} the melting criterion for a quasi two-dimensional superfluid is given as: \[
\frac{T_m}{T_{\rm BKT}}=\frac{1}{4\pi\sqrt{3}}\frac{n_s(T_m)}{n_s(T_{\rm BKT})}.
\]
This relation can be written in terms of superfluid phase-space densities: $\mathcal{D}_s(T_m)=4\pi\sqrt{3}\,\mathcal{D}_s(T_{\rm BKT})$. Using the known value for the superfluid jump at the transition $\mathcal{D}_s(T_{\rm BKT})=4$ \cite{Nelson1977}, we arrive at $\mathcal{D}_s(T_m)\simeq 87$. Evaluating precisely the superfluid phase-space density for a quasi two-dimensional quantum gas is not simple \cite{Holzmann2008a,Christodoulou2021,Chauveau2023}, however we may obtain an upper bound by using a zero-temperature Thomas-Fermi (TF) model as an ansatz for the density. Using a two-dimensional model for a harmonic plus quartic potential~\cite{Cozzini2005} we write:
\[
\rho(r)=\rho_0\left(1-\frac{r^2}{R_+^2}\right)\left(1-\frac{r^2}{R_-^2}\right),
\]
where $\rho_0=\mu\times M/(\hbar^2\tilde{g})$ and
\[
R_\pm^2=\frac{a_r^2}{2\kappa}\left[\frac{\Omega^2}{\omega_r^2}-1\pm\sqrt{\left(\frac{\Omega^2}{\omega_r^2}-1\right)^2+\frac{8\kappa\mu}{\hbar\omega_r}}\,\right].
\]
Here, $\mu$ is the chemical potential and $\tilde{g}=\sqrt{8\pi}a_s/a_z$ the dimensionless two-dimensional interaction constant \cite{Bloch2008RMP}, $a_s$ the scattering length and $a_z=\sqrt{\hbar/M\omega_z}$ the size of the vertical harmonic ground state. We consider only the case $\Omega<\omega_r$ such that $R_-^2<0$ while $R_+^2>0$ always holds.
The normalization $N=2\pi\int_0^{R_+}rdr\,\rho(r)$ fixes the chemical potential $\mu$ which in turns gives the peak density $\rho_0$. We finally obtain $T_m\leq 2\pi\hbar^2\rho_0/[Mk_B\mathcal{D}_s(T_m)]$, from which we evaluate $T_m$ numerically.

We may also compare this value with a simpler model based on a two-dimensional harmonic trap, for which analytical formulas for a Bose-Einstein condensate (BEC) exist. In this case the 
TF chemical potential is given by $\mu=\hbar\omega_r[\sqrt{8/\pi}\times N_0a_s/a_z\times(1-\Omega^2/\omega_r^2)]^{1/2}$, where $N_0$ is the condensed atom number. The phase-space density then reads:
\[
\mathcal{D}_s=2\pi\frac{T_c^0}{T}\sqrt{\frac{\pi f_c}{6\tilde{g}}}
\]
where $f_c=N_0/N$ is the condensed fraction and $T_c^0=\hbar\omega_r/(\pi k_B)\times[6N(1-\Omega^2/\omega_r^2)]^{1/2}$ is the two-dimensional ideal gas BEC critical temperature. Evaluating the phase-space density at the melting temperature gives: $T_m\simeq0.24 T_c^0$, where we have taken into account the small correction coming from the condensed fraction: $f_c(T_m)=1-(T_m/T_c^0)^2$, for a 2D harmonic trap.

\begin{figure}[t]
\centering
\includegraphics[width=8.6cm]{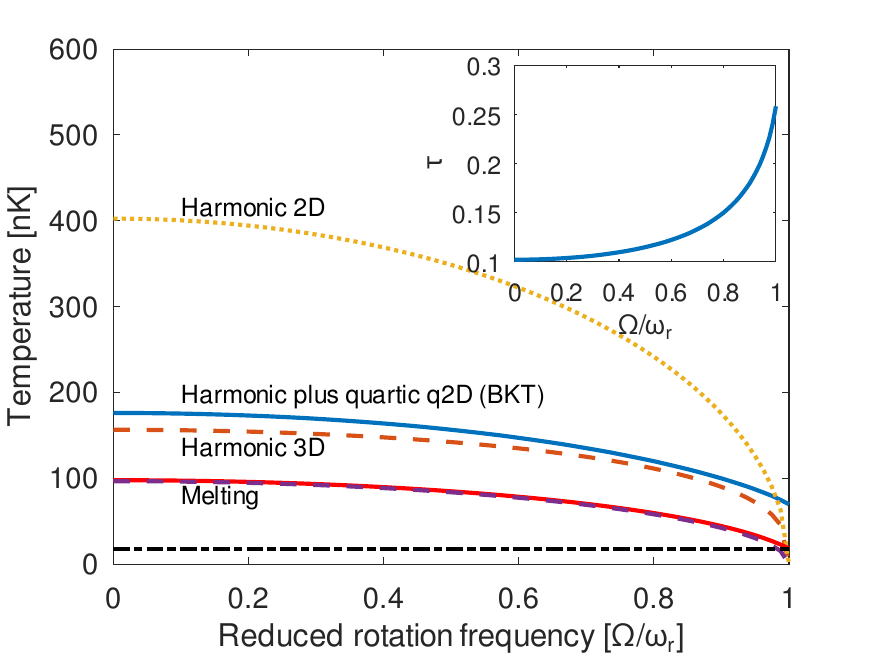}
\caption{
(color online) Estimation of the critical temperatures for superfluidity with $N=10^5$ atoms according to different models: the semi-classical BKT quasi-2D model for a harmonic plus quartic trap (solid blue curve), a 2D ideal gas in a harmonic trap (dotted yellow curve) and a 3D ideal gas in a harmonic trap (dashed orange curve). The two solid red and dashed violet curves labelled as `Melting' indicate the melting temperature estimated for the semi-classical BKT and ideal 2D models, respectively. The horizontal dashed-dotted black line indicates the temperature measured in the experiments. Inset: reduced temperature $\tau=T/T_{\rm BKT}$ as a function of the reduced rotation frequency (solid blue line), with $T=\SI{18}{nK}$.
}
\label{fig:Tmelting}
\end{figure}

Figure~\ref{fig:Tmelting} shows the values of the critical and melting temperatures estimated using the different models discussed above, for our experimental parameters. Concerning the melting temperature, there is an excellent agreement between the numerical estimation using the full semi-classical BKT model~\cite{Holzmann2008}, including the quartic correction, and the simple analytical formula for the 2D harmonic oscillator model. For this quantity, the effect of the quartic correction is important only for rotation frequencies $\Omega>0.9\omega_r$. As mentioned in the main text, the estimation of the melting temperature is much higher than the measured one.

Figure~\ref{fig:Tmelting} also suggests that performing experiments at fixed rotation frequency and atom number by varying the temperature implies to take into account fully the dimensional crossover, for our trap geometry. Indeed the semi-classical BKT temperature is closer to the 3D ideal gas prediction whereas the superfluid (or condensate) density is essentially confined to the ground state of the harmonic confinement to the surface.

\section{Classical field simulations}
We use a simple growth stochastic projected Gross-Pitaevskii equation (SPGPE) to sample the finite temperature equilibrium state of the system in a rotating frame~\cite{Bradley2007}, evolving a classical field $\psi_\mathcal{C}$ from a vacuum state, according to:
\[
i\hbar\frac{\partial\psi_\mathcal{C}}{\partial t}=\mathcal{P}_\mathcal{C}\left[(1-i\gamma)\left(H_0+g|\psi_\mathcal{C}|^2-\mu-\Omega L_z\right)\psi_\mathcal{C}+\eta\right],
\]
where $H_0=-\hbar^2/(2M)\times\bm{\nabla}^2+V(r,z)$ is the single particle Hamiltonian with $V(r,z)$ given at Eq.~\eqref{eq:potential}, and the fluctuating field $\eta$ is characterized by its second order correlations: $\braket{\eta(\bm{r},t)^*\eta(\bm{r}^\prime,t^\prime)}=2\hbar k_BT\gamma\delta(\bm{r}-\bm{r}^\prime)\delta(t-t^\prime)$. The approximate potential is separable, which enables an efficient implementation of the projector $\mathcal{P}_\mathcal{C}$ using a mixed Laguerre-Gauss / Gauss-Hermite basis along $r$ and $z$ respectively and an exact evaluation of the non-linear term with the appropriate quadrature~\cite{Bradley2007}.

We then evaluate the equilibrium state on a regular three dimensional grid in real space and use a split-step Fourier based numerical scheme to solve the Gross-Pitaevskii equation (GPE) describing the free expansion, using scaled coordinates: $\tilde{\bm{r}}\equiv(x/\lambda_r(t),y/\lambda_r(t),z/\lambda_z(t))$, where $\lambda_r(t)=\sqrt{1+(\Omega t)^2}$ takes into account a ballistic scaling due to the rotation and $\lambda_z(t)=\sqrt{1+(\omega_zt)^2}$ reflects the fast vertical expansion. This allows to track the full dynamics on a fixed grid of size $512\times521\times16$, despite the large magnification of all scales during the expansion.

\begin{figure}[t]
\centering
\includegraphics[width=8.6cm]{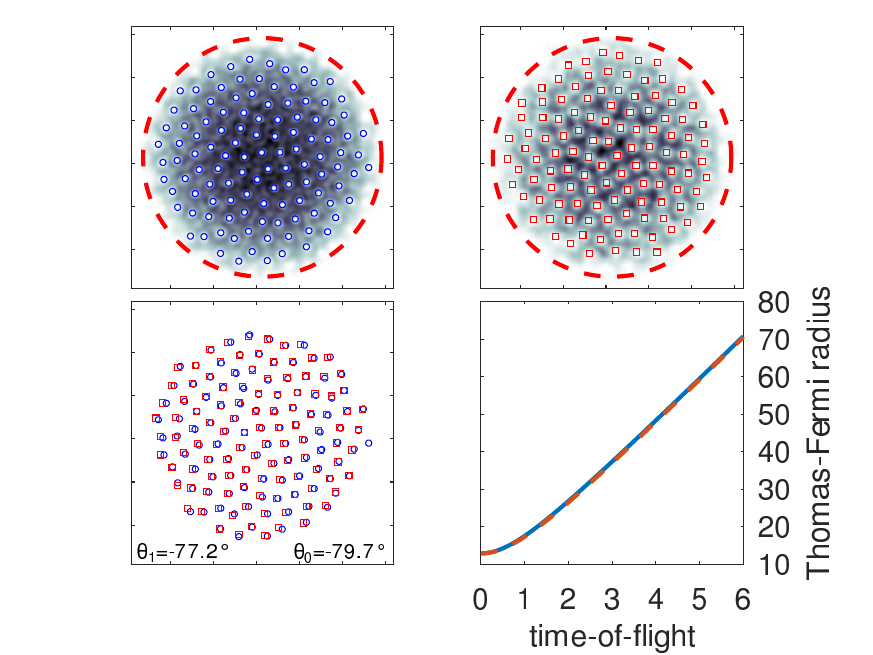}
\caption{
(color online) a) Density computed using a SPGPE simulation of the equilibrium state for $N=10^5$ atoms, a harmonic plus quartic trap with $\kappa=1.5\times10^{-4}$, $\Omega/\omega_r=0.9$ and $T=\SI{18}{nK}$. b) Same density after a \SI{28}{ms} time-of-flight expansion. The dashed red circles identify the TF radius. The length are scaled by $\sqrt{1+(\Omega t_{\rm tof})^2}$. In both pictures the positions of the vortices are identified with the blue circles and red squares, respectively. c) Comparison between the two vortex lattices, after scaling by $R_{\rm TF}^{\rm tof}/R_{\rm TF}^{\rm in\,situ}$ and rotating by an optimal angle $\theta_1$ slightly smaller (in magnitude) than the naive expectation $\theta_0=-\arctan{(\sqrt{1+(\Omega t_{\rm tof})^2}\times R_{\rm TF}^{\rm tof}/R_{\rm TF}^{\rm in\,situ})}$. d) Evolution of the TF radius during the time of flight $t_{\rm tof}$ from the full GPE simulation (blue line) and comparison with the ballistic scaling $\sqrt{1+(\Omega t_{\rm tof})^2}$ (red dashed line).
}
\label{fig:tofscaling}
\end{figure}

\begin{figure}[t]
    \centering
    \includegraphics[width=8.6cm]{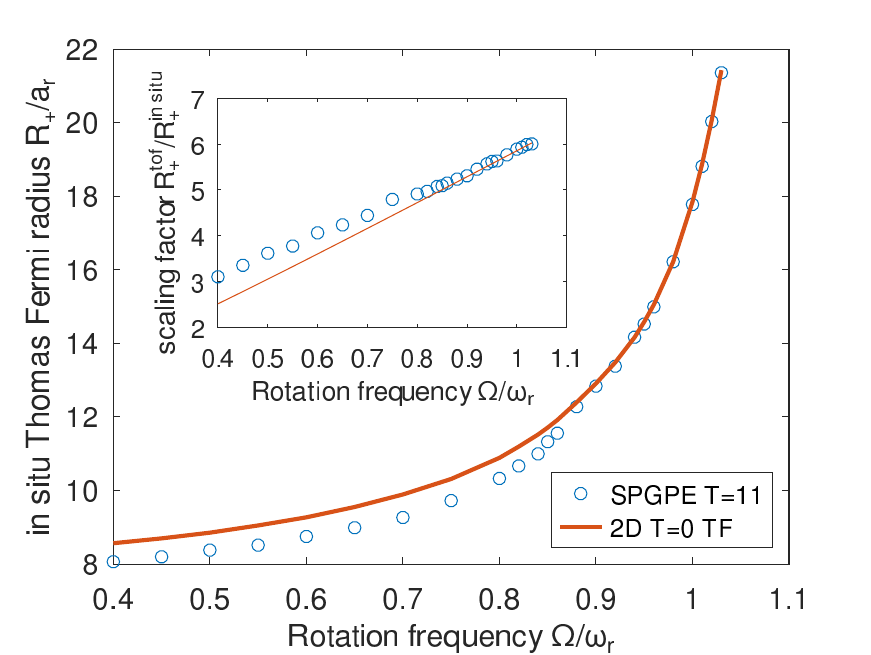}
    \caption{(color online) Benchmark of the SPGPE simulation: in situ TF radius in units of $a_r=\sqrt{\hbar/(M\omega_r)}$ as a function of the rotation frequency (symbols). Comparison with the two-dimensional $T=0$ TF solution. For $\Omega/\omega_r>0.85$ the SPGPE simulation is effectively 2D. Inset: scaling factor compared to the ballistic expansion. At low rotation frequency the 3D character results in a faster than ballistic expansion.}
    \label{fig:CheckModel}
\end{figure}

\section{Orientational correlations}
We present here the same data as in Fig.~3a) of the main text, using a log-log scale and comparing explicitly with a fit by an algebraically decaying function. Due to the finite size of our system we cannot plot $|G_6(r)|$ over more than a decade. The small number of vortices also impacts the dynamical range of the signal, that saturates obviously at long distances at a non-zero value, motivating our choice of a model $f_a(r)=C/r^\eta+D$. For $\Omega\leq0.85\times\omega_r$ the exponent is close to $\eta\simeq0.5$, significantly larger than the expected value of 0.25 at the hexatic to liquid transition.
\begin{figure}[t]
    \centering
    \includegraphics[width=8.6cm]{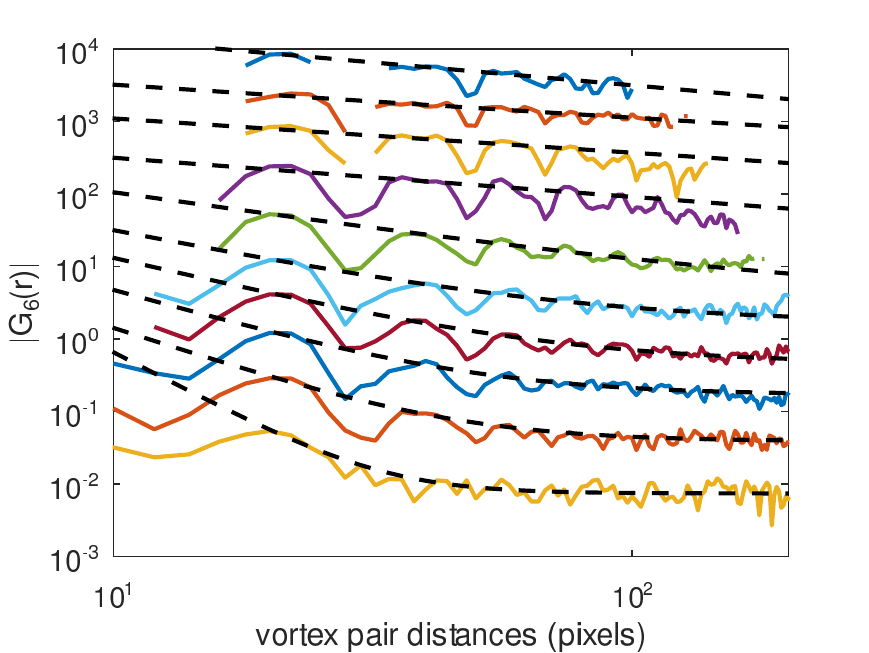}
    \caption{(color online) Decay of the orientational correlation function $|G_6(r)|$ with increasing reduced rotation frequency $\Omega/\omega_r=\{0.70,0.75,0.80,0.85,0.875,0.90,0.925,0.95,0.975,1.0\}$, from top to bottom, shown in log-log scale (same data as in Fig.~3a) 
    of the main text). The dashed lines show a fit by an algebraically decaying envelope, see text for details. The curves are shifted vertically for clarity.}
    \label{fig:G6algebraic}
\end{figure}

\section{Fit of the decaying envelope}
We detail the fitting procedure we use to estimate the characteristic decay length of the $|G_6(r)|$ correlations in Fig.~3a) 
of the main text. The difficulty is that we do not have a trial function that correctly reproduces the data for all dataset: at low rotation frequency the measured $|G_6(r)|$ shows large irregular oscillations, while for higher rotation frequency the behavior tends towards a damped cosine oscillation. We want to extract the characteristic decay length of the upper envelop of $|G_6(r)|$, which amounts to fit the curve only using its local maxima. To achieve this goal we applied two different methods that give similar results.

The first one relies on the output of the analysis of the pair correlation function $g(r)$, see Fig.~2 of the main text. Indeed, for a given rotation frequency, the position of the local maxima of $g(r)$ roughly coincide with the local maxima of $|G_6(r)|$. By selecting only the data points close to the local maxima (in a range $\pm 3$ pixels), we obtain an estimate of the envelope that we fit with an exponential decay.
The values reported in Fig.~3)b 
of the main text, as well as the graphs of Fig.~3a) 
and Fig.~\ref{fig:G6algebraic} are obtained with this first fitting method.

The second one relies on an iterative process: for a given rotation frequency we fit the $|G_6(r)|$ measurement by a decaying exponential. We then compare the measured data to the fitting curve and filter out all data points that are below the model by more than one sigma statistical uncertainty. We then repeat the fit taking into account the remaining points and iterate the procedure. After a few steps the procedure converges and reproduces the upper envelope of the measurements. This second method gives similar results and serves as a benchmark of the first method.

\end{document}